%% file: draft_clean.tex
\documentclass[fleqn,usenatbib]{mnras}
\usepackage{newtxtext,newtxmath}

\usepackage[T1]{fontenc}
\usepackage{ae,aecompl}

\usepackage{lastpage, graphicx, amssymb, aas_macros, bm, nicefrac}
\usepackage{bigints}
\usepackage{tabularx}
\usepackage{colortbl}
\usepackage[dvipsnames]{xcolor}
\hypersetup{breaklinks, colorlinks=true, urlcolor=blue, citecolor=blue, linkcolor=black}
\newcommand{\fbary}{f_{\rm{b}}}
\newcommand{\cbary}{c_{\rm{b}}}
\newcommand{\mbary}{M_{\rm{b}}}

\newcommand{\gvir}{g_{\rm vir}}

\newcommand{\gmax}{g_{\rm max}}
\newcommand{\sigcrit}{\Sigma_{\rm crit}}
\newcommand{\gcrit}{g_{\rm crit}}
\newcommand{\rtilde}{\tilde{r}}
\newcommand{\mss}{{\rm m\,s^{-2}}}
\newcommand{\rcrit}{r_{\rm crit}}
\newcommand{\mtotcrit}{M_{\rm tot,crit}}
\newcommand{\tff}{t_{\rm ff}}
\newcommand{\epsff}{\epsilon_{\rm ff}}
\newcommand{\epsstar}{\epsilon_{\star}}
\newcommand{\etaff}{\eta_{\rm ff}}
\newcommand{\mspsq}{M_{\odot}\,{\rm pc}^{-2}}

\definecolor{Gray}{gray}{0.9}
\definecolor{texas}{HTML}{BF5700}
\definecolor{update}{HTML}{800000}

\defcitealias{dekel2023}{D23}

\input{extra_preamble.tex}

\title[Efficient dark-matter-driven $z \gtrsim \mathit{8}$ galaxy formation]
{
Accelerated by dark matter: a high-redshift pathway to efficient galaxy-scale star formation
}
\author[M.~Boylan-Kolchin]
{Michael Boylan-Kolchin$^1$\thanks{$\!$mbk@astro.as.utexas.edu}\\
\noindent $\!\!^1$Department of Astronomy, The University of Texas at Austin,
2515 Speedway, Stop C1400, Austin, TX 78712-1205, USA}
\date{Accepted 2025 March 20. Received 2025 March 18; in original form 2024 July
15}

\pubyear{2025}

\begin{document}
\label{firstpage}
\pagerange{\pageref{firstpage}--\pageref{LastPage}}
\maketitle

\begin{abstract}
  In the local Universe, star formation is typically inefficient both globally
  and when considered as the fraction of gas converted into stars per local
  free-fall time. An important exception to this inefficiency is regions of high
  gravitational accelerations $g$, or equivalently surface densities
  $\Sigma = g/(\pi\,G)$, where stellar feedback is insufficient to overcome the
  self-gravity of dense gas clouds. In this paper, I explore whether dark matter
  can play an analogous role in providing the requisite accelerations on the
  scale of entire galaxies in the early cosmos. The key insight is that
  characteristic accelerations in dark matter halos scale as $(1+z)^2$ at fixed
  halo mass. I show this is sufficient to make dark matter the source of intense
  accelerations that might induce efficient star formation on galactic scales at
  cosmic dawn in sufficiently massive halos. The mass characterizing this regime
  scales as $(1+z)^{-6}$ and corresponds to a relatively constant comoving
  number density of $n(>\!\mvir) \approx 10^{-4}\,\mpc^{-3}$ at $z \gtrsim
  8$. For somewhat rarer halos, this model predicts stellar masses of
  $\mstar \sim 10^{9}\,\msun$ can form in regions that end up with sizes
  $\mathcal{O}(100\,\pc)$ over $40\,{\rm Myr}$ time-scales at $z\approx 12-14$;
  these numbers compare well to measurements for some of the brightest galaxies
  at that epoch from James Webb Space Telescope (JWST) observations. Dark matter
  and standard cosmological evolution may therefore be crucial for explaining
  the surprisingly high levels of star formation in the early Universe revealed
  by JWST.
\end{abstract}

\begin{keywords}
galaxies: formation -- galaxies: high-redshift -- dark matter -- cosmology: theory
\end{keywords}

\section{Introduction} 
\label{sec:intro}

Star formation is generally regulated by stellar feedback: young, massive stars
have prodigious UV output, leading to a variety of physical mechanisms that
inhibit further star formation. The star formation efficiency $\epsff$ --- the
fraction of gas converted into stars on a free-fall time --- is therefore low,
typically $\lesssim 2\%$, even in molecular clouds \citep{kennicutt1998a,
  murray2010, krumholz2019, hu2022, evans2022}. An important exception is dense
regions where baryons experience high accelerations: in this case, momentum
injection from massive stars, $\langle \dot{p}/m_{\star} \rangle$, is
insufficient to overcome gravity and star formation becomes efficient:
$\mstar=\epsstar\, M_{\rm gas}$, with $\epsstar\sim
\mathcal{O}(1)$. Observational, theoretical, and numerical results
\citep{fall2010, colin2013, geen2017, kim2018b, kruijssen2019c, grudic2020,
  polak2023} all point to a critical acceleration of
$\gcrit \approx \langle \dot{p}/m_{\star} \rangle \approx 5\times
10^{-10}\,\mss$ separating the regimes where stellar feedback removes most of
the potentially star-forming gas ($g \ll \gcrit$) and where gravity overcomes
the effects of feedback ($g \gg \gcrit$). For historical reasons, this is
usually expressed in terms of a surface mass density\footnote{It is therefore
  sometimes useful to express $G$ in relevant units as \\
  $1.39\times10^{-13}\,(\mspsq)^{-1}\,\mss$ or
  $4.50 \times 10^{-3}\,(\mspsq)^{-1}\,\pc\,{\rm Myr}^{-2}$},
$\Sigma\equiv M/(\pi\,R^2)=g/(\pi \,G)$; in these terms, the critical value is
$\sigcrit \approx 1000\,\mspsq$.
\begin{figure*}
 \centering
 \includegraphics[width=0.9925\columnwidth]{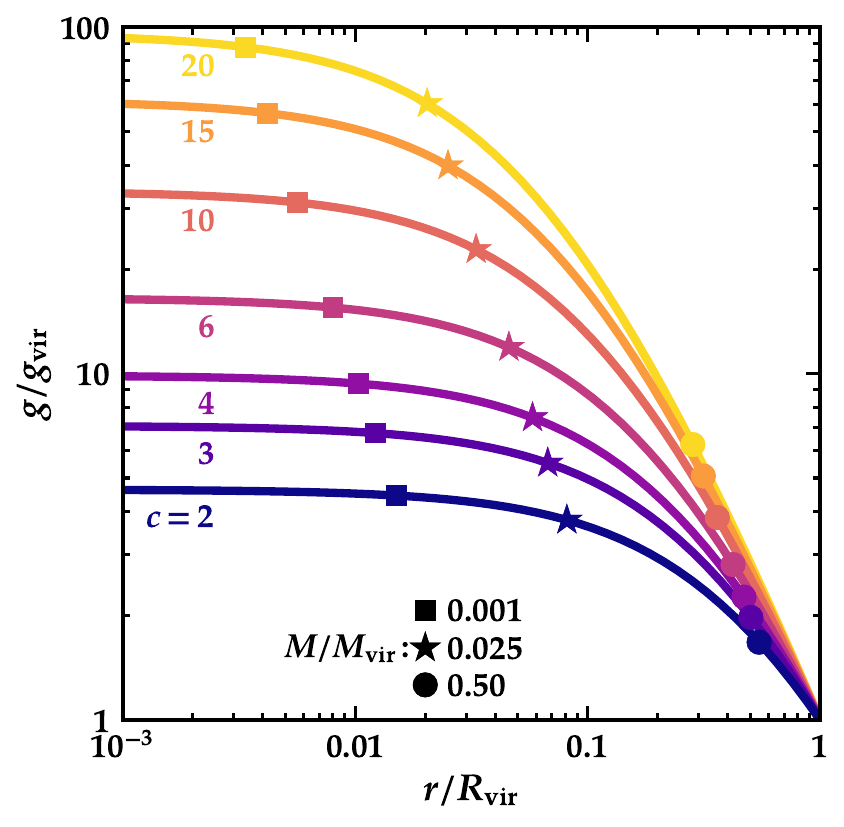}
 \includegraphics[width=1.0225\columnwidth]{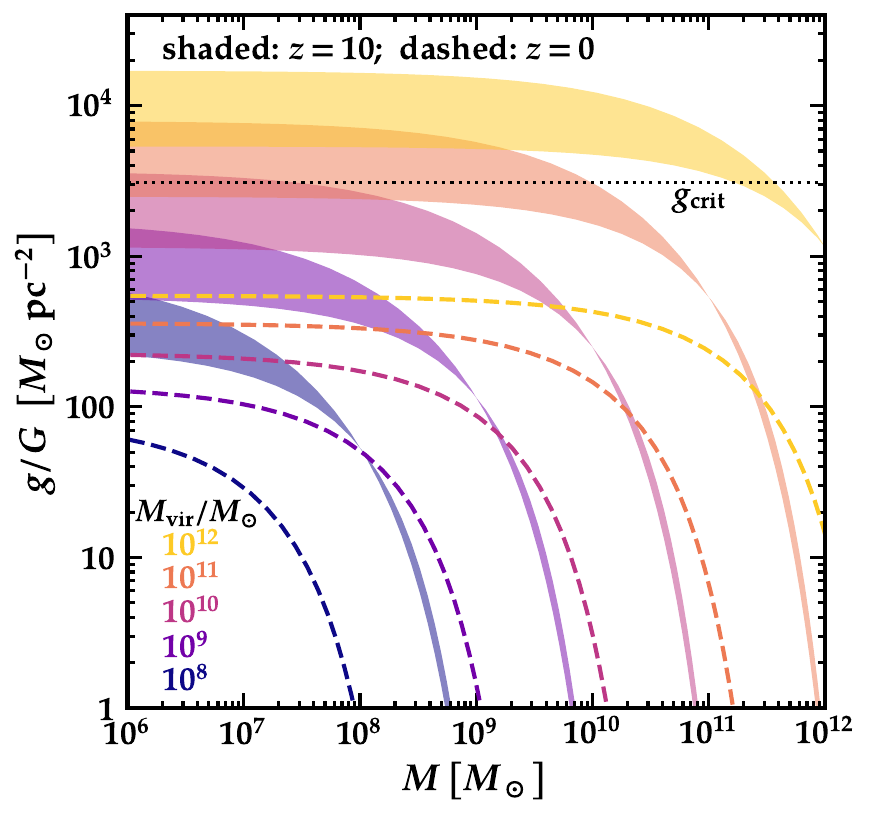}
 \caption{\textit{Left}: the acceleration profile of NFW halos (colored curves,
   with concentration labeled) as a function of radius, in virial units. The
   corresponding enclosed mass, also in virial units, is indicated on the
   plot. $\ssim 30\%$ or more of the mass in very concentrated halos lies above
   $10\,\gvir$ (and resides within $\ssim 20\%$ of $\rvir$). In
   low-concentration halos, the maximum acceleration of $\ssim (5\;{\rm
     to\;}7)\,\gvir$ is reached at $\sim 2\%$ of the virial radius; this region
   contains $\lesssim 1\%$ of such halos' mass. 
 \textit{Right}: acceleration profiles (scaled by $G$), now plotted as a
 function of enclosed mass $M(<r)$, of NFW halos with
 $\mvir/\msun=10^{8,\,9,\,10,\,11,\,{\rm and}\,12}$ at $z=0$ (dashed lines) and
 $z=10$ (shaded regions). I assume the median $c(\mvir|z=0)$ relation from
 \citet{ishiyama2021}, while the $z=10$ curves show the $1\,\sigma$ region
 around the mean from \citet{yung2024a}. The value of $\gcrit/G=3100\,\mspsq$
 adopted throughout this work is shown as a horizontal dotted line. At $z=0$, $g
 \ll \gcrit$ for all of the halo masses plotted; at $z=10$, a substantial
 fraction of the total mass in massive halos can lie above $\gcrit$. 
 \label{fig:nfw_g_profile}
}
\end{figure*}

In principle, this acceleration can be provided by any kind of matter. In
practice, regions in the local Universe where large amounts of gas experience
high enough accelerations to undergo efficient star formation are dense
concentrations of baryons, typically in the form of molecular clouds. The vast
majority of dark matter must be effectively dissipationless on scales relevant
for galaxy formation, meaning that it is unable to cool to high enough densities
to contribute to high accelerations given the measured cosmological mean value
of the dark matter density at $z=0$. As I discuss in more detail in
Sec.~\ref{sec:background}, the acceleration at the outer edge of a galaxy-scale
dark matter halo at $z=0$ is $g/G \approx (14\,\mspsq)$ while the acceleration
at its center is typically $\sim 30$ times higher ($\approx 400\,\mspsq$), still
well below $\gcrit$.

However, at fixed halo mass, the acceleration at the virial radius scales
roughly as $(1+z)^{2}$, which means halos of a fixed mass are subject to
accelerations $\ssim 100$ times higher at $z \approx 10$ than at
$z=0$. \textit{Dark matter} may then provide the high accelerations needed for
efficient star formation. As I demonstrate below, the potential wells of dark
matter halos at redshifts $\gtrsim 8$ can subject large masses of baryons to
these high accelerations, providing a potential avenue for efficient star
formation on a galaxy-wide scale.

Galaxy formation efficiency at these redshifts has recently shifted from purely
theoretical speculation to an urgent observational and theoretical puzzle. JWST
has revealed an epoch of strikingly and unexpectedly active galaxy and black
hole growth at cosmic dawn (\citealt{eisenstein2023, greene2023, akins2024,
  donnan2024, dressler2024}; for a recent review, see \citealt{adamo2024}), and
a variety of models have been used to interrogate \citep{boylan-kolchin2023,
  keller2023, lovell2023, mason2023, shen2023} or explain \citep{dekel2023,
  ferrara2023, mirocha2023, mcgaugh2024, nusser2024, rennehan2024} these
surprising results. The goal of this paper is to point out that efficient star
formation on large scales --- significantly exceeding the mass scales of giant
molecular clouds --- is a simple but unavoidable consequence of the evolution of
dark matter densities in an expanding Universe, which produce much more intense
gravitational accelerations in galaxy-scale halos at high redshift than are
possible in the local Universe, and to explore some of the attendant
implications for galaxy formation.

When necessary, I assume a standard dark energy + dark matter \lcdm\ cosmology with
$H_0=67.32\,\kms\,\mpc^{-1}$, $\Omega_{\rm m}=1-\Omega_{\Lambda}=0.3158$,
$n_{\rm s}=0.96605$, $\sigma_{8}=0.8120$, and
$\fbary \equiv \Omega_{\rm b}/\Omega_{\rm m}=0.156$~\citep{planck2020}. I adopt
$\sigcrit=1000\,\mspsq \leftrightarrow \gcrit/G = 3100\,\mspsq$ for
concreteness. The precise value of $\gcrit$ does not matter for the qualitative
picture I describe, as the relationship between acceleration or total surface
density and integrated star formation efficiency increases quickly for
$g \ll \gcrit$ and saturates for $g \gtrsim \gcrit$ \citep{fall2010,
  grudic2020}. Quantitative predictions will be sensitive to the precise value
of $\gcrit$, a point to which I return in Sec.~\ref{sec:speculation}.

\section{Background}
\label{sec:background}
The virial radius of a dark matter halo of mass $\mvir$ at redshift $z$ is defined via 
\begin{equation}
    \label{eq:vir_def}
    \mvir=\frac{4\,\pi}{3}\,\rvir(z)^{3}\,\Delta(z)\,\rho_{\rm m}(z)\,,
\end{equation}
where $\Delta_{\rm vir}(z)\equiv \Delta(z)\,\Omega_{\rm m}(z)$ is the
overdensity calculated using the spherical top-hat collapse model
\citep{bryan1998}. The acceleration at the virial radius --- the virial
acceleration $\gvir$ --- is then
\begin{equation}
\label{eq:gvir}
    \gvir \equiv \frac{G\,\mvir}{\rvir^2}\,.
\end{equation}

The acceleration profile of a dark matter halo interior to the virial radius
follows from its mass profile; for a \citet*[hereafter, NFW]{navarro1996,
  navarro1997} profile, the acceleration at radius $\rtilde\equiv r/\rvir$
depends only on the virial acceleration and the halo concentration:
\begin{equation}
\label{eq:gprofile}
 g(r)=\frac{\gvir}{\mu(c)} \frac{\mu(c\,\rtilde)}{\rtilde^2}
\end{equation}
where $\mu(x) \equiv \ln(1+x)-x/(1+x)$. As $r \rightarrow 0$, the density
profile is $\rho \propto r^{-1}$, giving a mass profile of $M(<r) \propto r^{2}$
and an acceleration profile that approaches a constant, maximum value:
\begin{equation}
\label{eq:gmax}
 \gmax=\frac{\gvir}{\mu(c)}\frac{c^2}{2}\,
\end{equation}
(e.g., \citealt{power2003, navarro2017}).

The left panel of Fig.~\ref{fig:nfw_g_profile} shows the acceleration profile
for NFW halos with a variety of concentrations ranging from $c=2$ to $20$
plotted as a function of $r/\rvir$. Symbols mark the indicated fixed fractions
of enclosed mass relative to $\mvir$. For very concentrated halos, 50\% of the
mass experiences an acceleration that is greater than $7\,\gvir$, extending over
$30\%$ of the halo, while none of the mass in $c=2$ halos experiences an
acceleration exceeding $5\,\gvir$.

\section{Accelerations at high redshift}
\label{sec:highz}
\subsection{Relating halo mass, virial accelerations, and halo abundances across redshifts} 
\label{subsec:z_dep}
Within a halo's virial radius, the range of accelerations experienced due to
dark matter alone is therefore relatively narrow, a factor of $10$ for a
fiducial concentration of $c=4$. The Milky Way ($\mvir=10^{12}\,\msun$,
$c\approx 10$) has $\gvir(z=0)/G \approx 14\,\mspsq$, so the maximum
acceleration from dark matter is $\gmax/G \approx 500\,\mspsq$ or
$\gmax \approx 7 \times 10^{-11}\,\mss$, well below $\gcrit$. Even in
galaxy-cluster-mass halos with $\mvir(z=0) \approx 10^{15}\,\msun$, $\gvir/G$ is
only $140\,\mspsq$ and $\gmax/G \approx 1500\,\mspsq$. However, at fixed halo
mass, the virial radius scales inversely with $(1+z)$:
\begin{equation}
\label{eq:rvir}
\rvir=7\,\kpc 
\left( \frac{\mvir}{10^{10}\,\msun}\right)^{1/3}\,\left(\frac{1+z}{10}\right)^{-1} \; \left(\frac{\Delta(z)}{{18\,\pi^2}}\right)^{-1/3}\,.
\end{equation}
The virial acceleration therefore increases as $(1+z)^2$:
\begin{flalign}
\label{eq:gvir_mss}
\gvir &=2.84 \times 10^{-11}\,\mss 
\left( \frac{\mvir}{10^{10}\,\msun}\right)^{1/3}\,\left(\frac{1+z}{10}\right)^{2} \; \left(\frac{\Delta(z)}{{18\,\pi^2}}\right)^{2/3}\\
\label{eq:gvir_mspsq}
\frac{\gvir}{G}&=204 \,\mspsq\,
\left( \frac{\mvir}{10^{10}\,\msun}\right)^{1/3}\,\left(\frac{1+z}{10}\right)^{2} \; \left(\frac{\Delta(z)}{{18\,\pi^2}}\right)^{2/3}\,.
\end{flalign}
The centers of galaxy-scale halos with $\mvir \sim 10^{10}-10^{12}\,\msun$ can
therefore reach or even exceed $\gcrit$ at high redshifts; even
$\mvir \approx 10^{9}\,\msun$ halos can have $g>\gcrit$ at $z \gtrsim 17$

To further emphasize this point, we can invert Eq.~\ref{eq:gvir_mspsq} to obtain
the virial mass as a function of the virial acceleration (or surface density)
and redshift:
\begin{equation}
    \label{eq:mvir_gvir}
    \mvir=10^{10}\,\msun\,\left( \frac{\gvir/G}{204\,\mspsq}\right)^{3}\,\left(\frac{1+z}{10}\right)^{-6}\,\left(\frac{\Delta(z)}{{18\,\pi^2}}\right)^{-2}.
\end{equation}
\textit{The virial mass resulting in a fixed virial acceleration scales as $(1+z)^{-6}$.}

This point is emphasized in the right panel of Fig.~\ref{fig:nfw_g_profile}. It
shows the relationship between enclosed mass and acceleration at $z=0$ (dashed
lines) and $z=10$ (shaded regions) for
$\log_{10}(\mvir/\msun)=8,\,9,\,10,\,11,\,{\rm and}\,12$. I assume the mean
$c(M | z=0)$ relation from \citet{ishiyama2021} as implemented in {\sc Colossus}
\citep{diemer2018}; at $z=10$, the shaded region corresponds to concentrations
between $c=2$ and $c=5.5$, which approximately spans the symmetric 68\% interval
around the median concentration found in the cosmological simulations of
\citet{yung2024a}. It is straightforward to see that virtually all of the mass
in a $\mvir(z=10)=10^{12}\,\msun$ halo lies above $\gcrit$, while a
$\mvir(z=10)=10^{11}\,\msun$ halo can have as much as $\sim 10^{10}\,\msun$ or
as little as no mass above $\gcrit$.

\begin{figure}
 \centering
 \includegraphics[width=\columnwidth]{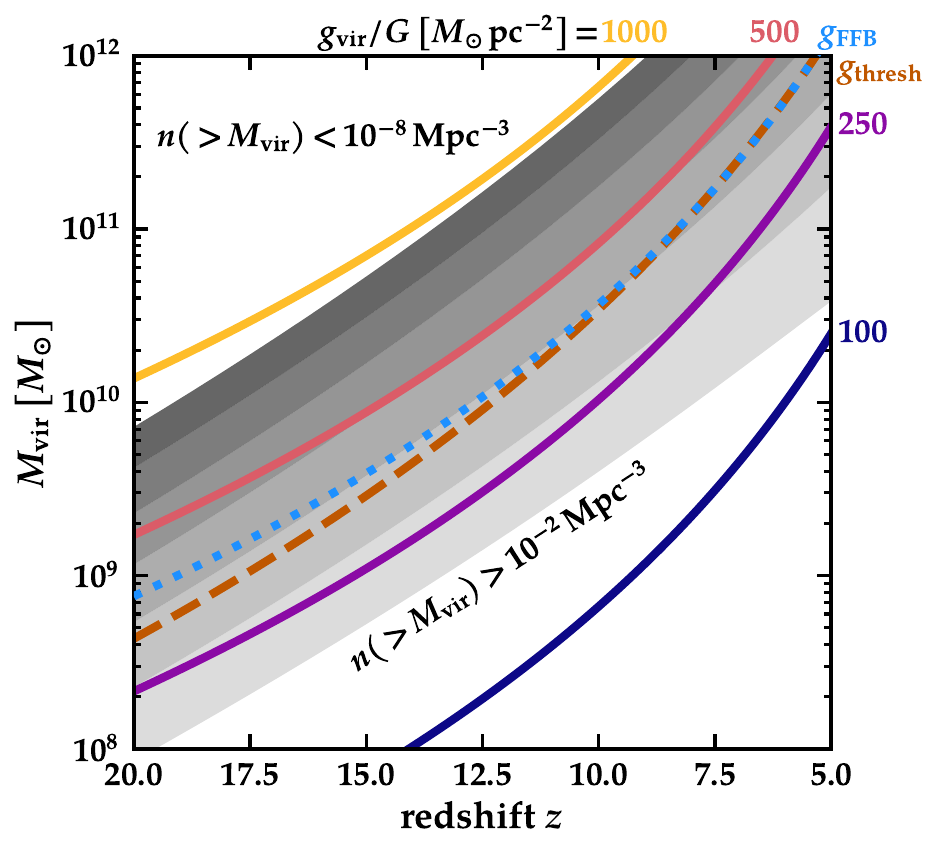}
 \caption{The solid, colored curves show the evolution of $\gvir/G=[100,\,250,
   \,500,\,{\rm and }\,1000]\,\mspsq$ halos in $\mvir-z$ space. The grayscale
   bands show the evolution of halos of fixed number densities, from
   $n(>\mvir)=10^{-8}\,\mpc^{-3}$ (top) to $n(>\mvir)=10^{-2}\,\mpc^{-3}$
   (bottom). Halos of fixed cumulative comoving number density closely track the
   evolution of halos with fixed values of $\gvir$ for the redshift range
   explored here. The dashed orange curve shows the $\mvir(z)$ evolution of
   $g_{\rm thresh}$ (where $\gmax(\mvir,z)=\gcrit$) explored in this paper,
   while the dotted blue curve shows $g_{\rm FFB}$ from
   \citetalias{dekel2023}. Intriguingly, $g_{\rm thresh}$ and $g_{\rm FFB}$
   nearly coincide for the full redshift range plotted, tracing out $n(>\mvir)
   \approx 10^{-4}\,\mpc^{-3}$. 
 \label{fig:m_vs_z}
}
\end{figure}
\begin{figure*}
 \centering
 \includegraphics[width=0.99\columnwidth]{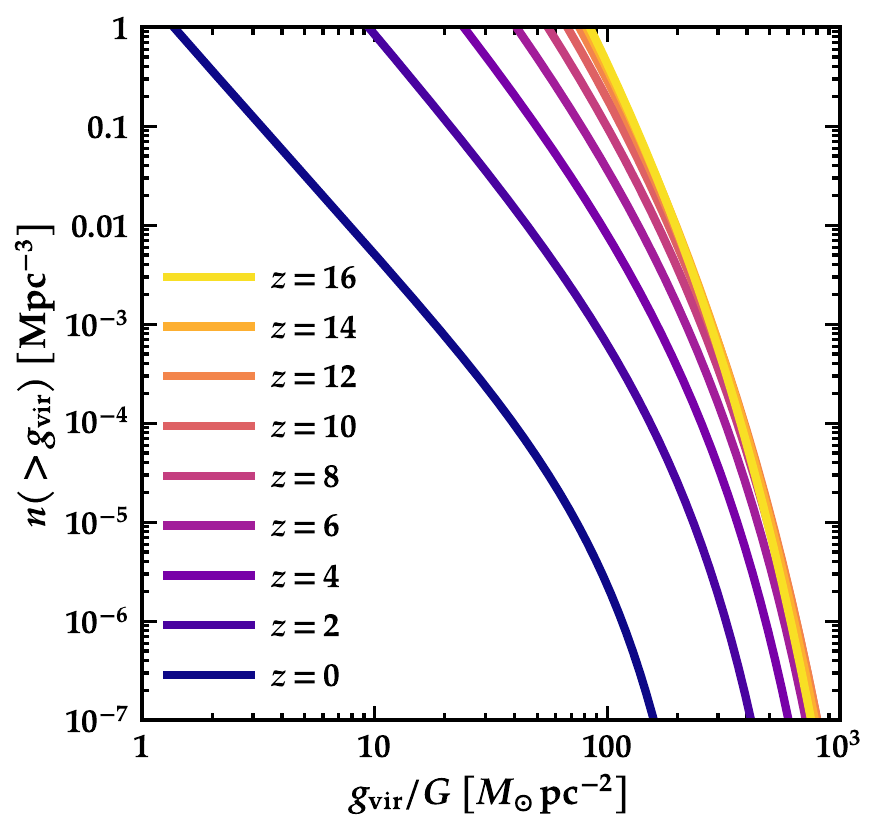}
 \includegraphics[width=1.01\columnwidth]{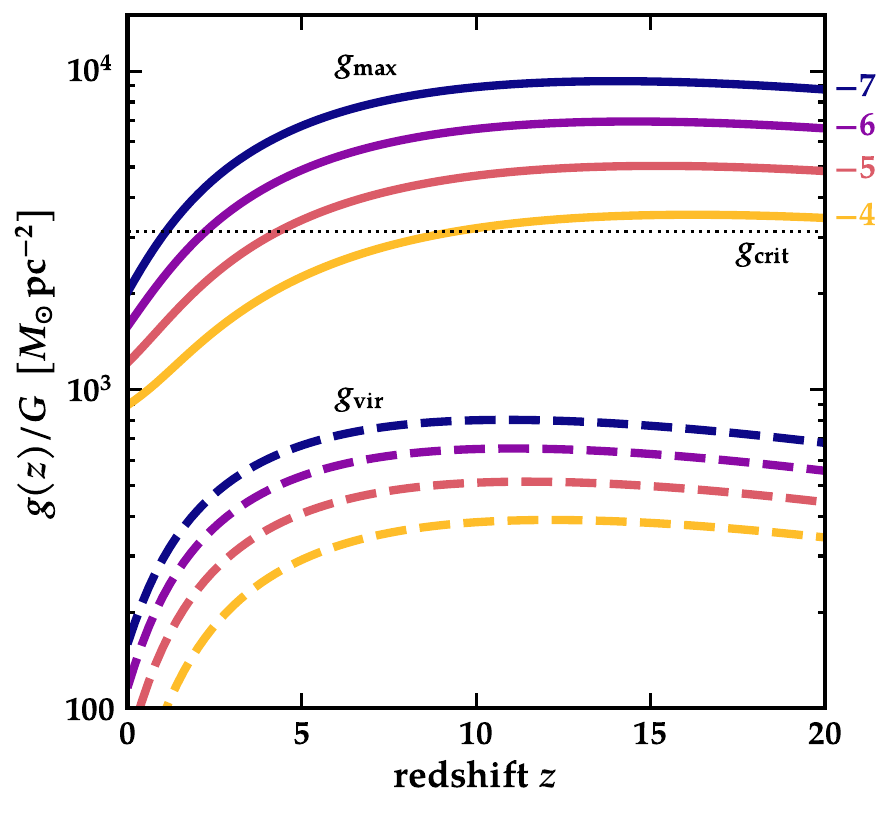}
 \caption{\textit{Left}: The cumulative $\gvir$ function of dark matter halos
   from $z=0$ (dark blue curve) to $z=16$ (yellow curve). $n(>\gvir)$ evolves
   rapidly from $z=0$ to $z \approx 6$, with the cumulative abundance above a
   fixed value of $\gvir$ increasing toward higher redshift, then changes very
   little at the highest redshifts. As a result, the number density of halos
   above a fixed $\gvir$ remains nearly constant above $z \sim
   6$. \textit{Right}: the redshift evolution of $\gvir$ (dashed lines) and
   $\gmax$ (solid lines, assuming the median $c(\mvir | z)$ from
   \citealt{yung2024a}) for fixed comoving number densities of
   $n(>\mvir)=10^{-(7,\,6,\,5,\,{\rm and }\,4)}\,\mpc^{-3}$. Both $\gmax$ and
   $\gvir$ at fixed number densities rise rapidly from $z=0$ to $z\approx 6$
   (for $\gvir$) or $z\approx 8$ (for $\gmax$), then remain virtually constant
   to $z=20$. All of the number densities plotted here have $\gmax > \gcrit$ at
   high redshift, meaning they are candidates for efficient dark-matter-driven
   galaxy-wide star formation.  
 \label{fig:gvir_func}
}
\end{figure*}

To understand whether dark matter can play a role in efficient star formation,
therefore, we must understand how likely it is to find halos with appreciable
mass above $\gcrit$ at high redshift. Figure~\ref{fig:m_vs_z} contains this
information: it shows the virial mass at redshift $z$ corresponding to
cumulative abundances ranging from $10^{-8}\,\mpc^{-3}$ (the upper boundary of
the darkest gray region) to $10^{-2}\,\mpc^{-3}$ (the lower boundary of lightest
gray region). It is immediately apparent that we cannot use
$\mvir=10^{12}\,\msun$ halos to explain efficient star formation at
$z \gtrsim 7$ revealed by JWST via high (dark matter) surface density: the
expected abundance of objects at least this massive is
$\approx 10^{-6.5}\,\mpc^{-3}$ at $z=7.5$ and drops precipitously as redshift
increases, while the observed number density of surprisingly bright galaxies at
high redshift is at least an order of magnitude larger.

The solid colored lines in Figure~\ref{fig:m_vs_z} show the evolution of halo
mass corresponding to fixed
$\gvir/G=(1000,\,500,\,250,\,{\rm and}\,100)\,\mspsq$. These curves are defined
by Eq.~\ref{eq:mvir_gvir} and therefore evolve as $(1+z)^{-6}$. Intriguingly,
they track the evolution of halos at fixed cumulative abundance quite closely
for $z \gtrsim 8$: \textit{halos of a fixed cumulative comoving number density
  have nearly fixed virial accelerations}. The dashed orange line shows a
threshold acceleration $g_{\rm thresh}$, defined via
$\gmax(\gvir=g_{\rm thresh},z) = \gcrit$, i.e., the virial acceleration where a
halo's central acceleration achieves $\gcrit$. This is the minimum requirement
for having a non-zero quantity of gas exceeding $\gcrit$ owing to accelerations
from high dark matter densities. This threshold is not a constant value with
redshift because halo concentrations at a given $\mvir$ evolve somewhat with
time, but the figure shows this evolution has a very minor effect on
$g_{\rm thresh}$: it closely follows the contour for a constant cumulative
comoving number density of $n \approx 10^{-4}\,\mpc^{-3}$ from
$z=20\,{\rm to}\,8$.

The blue dotted curve in Figure~\ref{fig:m_vs_z} shows the redshift evolution of
$\mvir$ giving $\gvir/G=381\,\mspsq$, which has
$\mvir(z=9)=10^{10.8}\,\msun$. This virial mass and redshift combination is
noteworthy because it was derived by \citet{dekel2023} for conditions conducive
to feedback-free bursts (FFBs) in the early Universe, where star formation is
postulated proceed in a highly efficient manner\footnote{I have ignored the
  slight difference in cosmology and virial mass definition adopted by
  \citetalias{dekel2023} in computing the virial acceleration for the
  characteristic FFB mass; this can lead to changes in the value of
  $g_{\rm FFB}$ at the few percent level, meaning the \textit{exact}
  correspondence seen at the lowest redshifts in Figure~\ref{fig:m_vs_z} is
  coincidental. In any case, the mass scale of $\mhalo=10^{10.8}\,\msun$ was not
  (and cannot be) characterized to percent-level accuracy in
  \citetalias{dekel2023}.}. The values of $g_{\rm FFB}$ and $g_{\rm thresh}$ and
their evolution with redshift are nearly identical, reinforcing the possibility
that \textit{efficient galaxy-wide star formation in the high-redshift Universe
  can be catalyzed by the gravity from dark matter in halos that exceed a
  threshold virial acceleration.}

An alternate way to look at the evolution of accelerations within halos at high
redshift is to plot the cumulative number density of halos as a function of
$\gvir(z)$ (recall that $\gvir$ is straightforwardly related to $\mvir(z)$ via
Eq.~\ref{eq:gvir_mss}). The left panel of Figure~\ref{fig:gvir_func} shows this
cumulative comoving number density as a function of $\gvir$ for redshifts from
$z=0$ to $z=16$. The number density at fixed $\gvir$ increases quickly from
$z=0$ to $z \approx 4-5$. At higher redshift, however, the evolution at fixed
$\gvir$ is almost negligible. Once again, we see that halos of a fixed number
density correspond very closely to halos of a fixed $\gvir$ at high redshift.

\begin{figure*}
 \centering
 \includegraphics[width=\columnwidth]{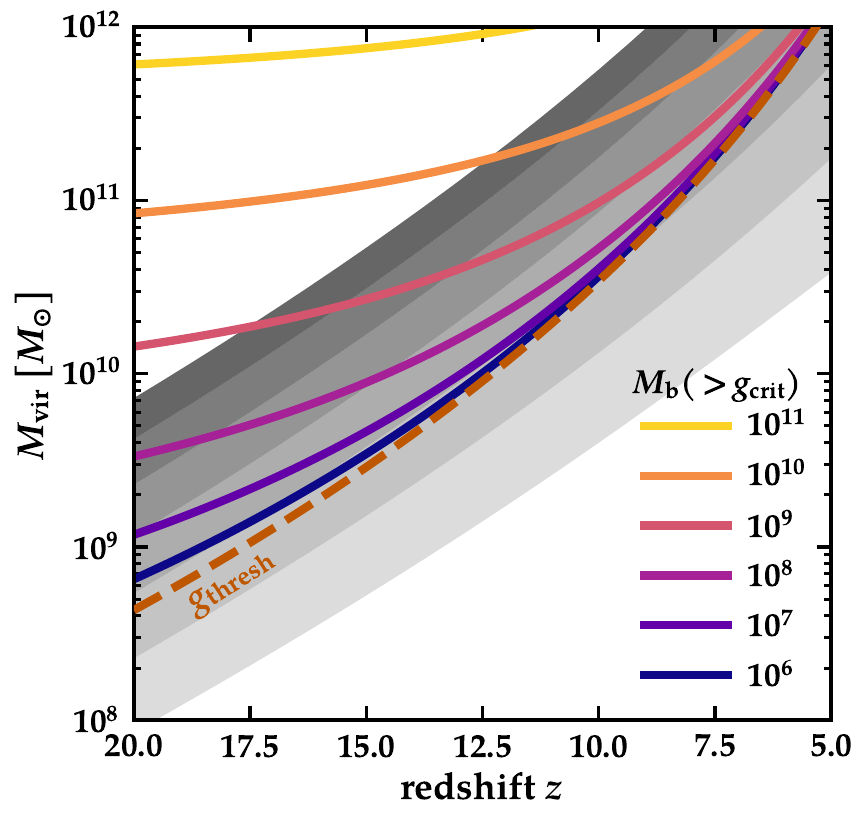}
 \includegraphics[width=\columnwidth]{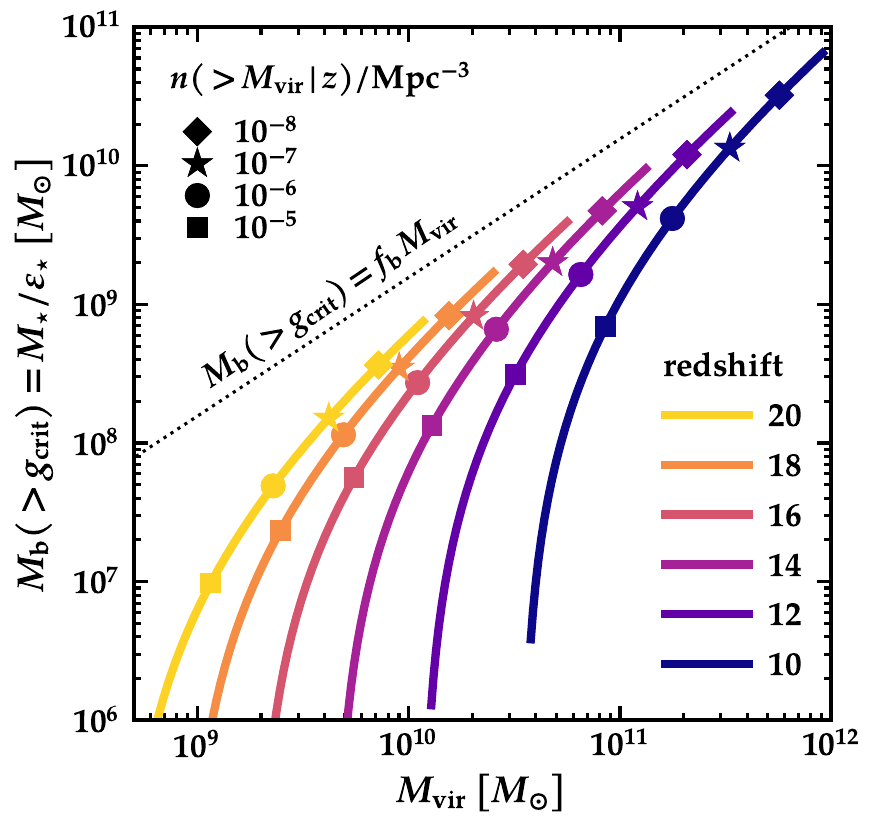}
 \caption{\textit{Left}: the solid, colored lines show the evolution of fixed
   masses in baryons that lie above $\gcrit$ as a function of $z$, from $M_{\rm
     b}(>\gcrit)=10^{11}\,\msun$ (yellow, top) to $10^{6}\,\msun$ (blue,
   bottom). As in Figure~\ref{fig:m_vs_z}, the gray shaded regions show the
   evolution of fixed cumulative comoving halo number densities (from
   $10^{-8}\,\mpc^{-3}$ at the top to $10^{-2}\,\mpc^{-3}$ at the bottom) and
   the dashed orange curve shows the evolution of $g_{\rm thresh}$ in the
   $(\mvir,z)$ plane. While it is essentially impossible to find halos massive
   enough to collect $10^{11}\,\msun$ of potentially star-forming material at
   high accelerations in the high-$z$ Universe, collections of $10^{10}\,\msun$
   ($10^{9}\,\msun$) are possible below $z \approx 10$ ($15$) and collections of
   $M_{\rm b}(>\gcrit)=10^{6}-10^{7}\,\msun$ are common even out to
   $z=20$. \textit{Right}: $M_{\rm b}(>\gcrit)$ as a function of $\mvir$ at
   several redshifts; the curves terminate on the right at
   $n(>\mvir)=10^{-9}\,\mpc^{-3}$. Fixed cumulative comoving number densities of
   halos are marked at each redshift with symbols. By redshift 20, halos
   containing $\mbary(>\gcrit) \approx 10^{7}\,\msun$ have number densities of
   $10^{-5}\,\mpc^{-3}$, while at $z=10$, the abundance of systems with
   $\mbary(>\gcrit) \approx 4\times10^9\,\msun$ ($\mstar \approx
   2\times10^9\,\msun$) is $\approx 10^{-6}\,\mpc^{-3}$. Note that at no point
   in time do halos reach the theoretical maximum of
   $\mbary(>\gcrit)=\fbary\,\mvir$.  
 \label{fig:m_z_baryon}
}
\end{figure*}

The right panel of Figure~\ref{fig:gvir_func} emphasizes this point further. The
dashed lines show the evolution of $\gvir(z)$ at fixed number densities of
$n(>\mvir)=10^{-(4,5,6,7)}\,\mpc^{-3}$, revealing the near constancy of $\gvir$
at fixed number density for $z \gtrsim 5$. The solid lines in the right panel of
Figure~\ref{fig:gvir_func} show how $\gmax$ evolves at the same cumulative
number densities; the $c(\mvir|z)$ relation adopted for this figure assume the
model of \citet{ishiyama2021} which predicts $c \approx 4.2 \pm 1.4$ at all
redshifts $\gtrsim 6$ for the number densities shown in the figure. The
abundance of halos as a function of $\gmax$ is also roughly constant in redshift
above $z \sim 7$. The value of $\gmax$ at high redshifts notably exceeds
$\gcrit$ for the four number densities plotted here. Dark matter halos at high
redshift may have sufficient internal accelerations to overcome momentum
injection from stellar feedback, resulting in efficient galaxy formation on
galaxy-wide scales. An additional important result from
Figure~\ref{fig:gvir_func} is that halos at a fixed number density lying above
$\gcrit$ have been above $\gcrit$ since at least $z\sim 20$, indicating the
efficient mode of star formation posited here is likely the first mode of star
formation these halos undergo: there is no chance for feedback from earlier
generations of star formation to lower the central densities (and therefore
central accelerations) of these halos.

\subsection{From halo properties to baryons}
\label{subsec:halos_to_baryons}
The remaining question, therefore, is how much mass \textit{in baryons}
experiences accelerations above $\gcrit$ at high redshifts? The left panel of
Figure~\ref{fig:m_z_baryon} shows the evolution with redshift of the
\textit{baryonic} mass above $\gcrit$; the gray-scale contours show the
evolution of the same fixed number densities as in Figure~\ref{fig:m_vs_z} while
the dashed orange line once again shows $g_{\rm thresh}$. For the purposes of
this plot, I assume that each halo has its cosmic fraction of baryons,
$M_{\rm b}(<\rvir)=f_{\rm b}\,\mvir$, and that the baryons have a spatial
distribution matching that of the dark matter.  This gives an upper limit to the
stellar mass content of a halo that can come from efficient conversion of
baryons in a dark matter halo via the mechanism described here. The plot
emphasizes the difficulty of getting efficient star formation at high redshift
even in the high acceleration regime. Halos with maximum accelerations just
reaching $g_{\rm thresh}$ have only $\sim 10^6\,\msun$ of baryons exceeding
$g_{\rm crit}$, independent of redshift. At $z=10$, the number density of halos
containing $10^{10}\,\msun$ in baryons above $\gcrit$ is
$\approx 10^{-7}\,\mpc^{-3}$; halos with $M_{\rm b}(>\gcrit)>10^{11}\,\msun$ are
rarer than $1\,{\rm Gpc}^{-3}$. Even with very efficient star formation in this
regime --- a conversion of all available baryonic mass above $\gcrit$ into
stars, meaning $\epsstar=1$ for this material --- halos with $10^{10}\,\msun$ of
stars will be very rare and halos with $\mstar \gtrsim 10^{11}$ should not exist
at $z \gtrsim 10$.

The right panel of Figure~\ref{fig:m_z_baryon} provides an alternate way to
understand the available baryon reservoirs for efficient star formation at high
redshift. It shows the mass in baryons above $\gcrit$ as a function of halo mass
at six different redshifts. The symbols mark cumulative comoving number
densities as noted on the plot. For surveys probing effective volumes of
$V\approx 10^5\,\mpc^3$, the rarest objects expected on average have
$n \approx V^{-1} \approx 10^{-5}\,\mpc^{-3}$, corresponding to the squares in
the figure. At $z \approx 20$, the most massive object in such a survey should
be no more than $\mstar \approx 10^{7}\,\msun$; by $z\sim 10$, objects with
$\mstar \approx 10^{9}\,\msun$ may be present. Surveys probing 100x larger
volumes will be able to see objects that are an order of magnitude more massive
at a fixed redshift. Note that even for extremely rare halos with
$n=1\,{\rm Gpc}^{-3}$, the baryon content at high accelerations does not
approach the theoretical maximum of $\fbary\,\mvir$, underlining the extreme
difficulty of converting anything close to a halo's cosmic fraction of baryons
into stars at high redshift.

\subsection{Star formation rates and stellar masses}
\label{subsec:sfr_and_mstar}
The region of efficient star formation has a size $\rcrit$ that is defined by
$g(<\rcrit)=\gcrit\equiv \mtotcrit/\rcrit^2$. For my choice of
$\gcrit/G=3100\,\mspsq$, the size, density and free-fall time in this region
are:
\begin{equation}
    \label{eq:rcrit}
    \rcrit=1.8\,\kpc\,\left(\frac{\mtotcrit}{10^{10}\,\msun} \right)^{1/2}\,
\end{equation}
\begin{equation}
    \label{eq:rhocrit}
    \langle \rho_{\rm tot}(<\rcrit) \rangle =0.4\,\msun\,\pc^{-3}\,\left(\frac{\mtotcrit}{10^{10}\,\msun} \right)^{-1/2},\;{\rm and}
\end{equation}
\begin{equation}
    \label{eq:tcrit}
    \tff(\rcrit)=\frac{\pi}{2}\sqrt{\frac{\rcrit^3}{2\,G\,\mtotcrit}}
    =12.7\,{\rm Myr}\,\left(\frac{\mtotcrit}{10^{10}\,\msun} \right)^{1/4}\,,
\end{equation}
where I have evaluated the free-fall time $\tff$ at the mean total density
$\langle \rho_{\rm tot}(<\rcrit) \rangle$. The baryonic mass available within
this radius is $\mbary(<\rcrit)=\fbary\,\cbary\,M_{\rm tot}(<\rcrit)$, where
$\cbary$ parameterizes the concentration of the baryons relative to the dark
matter within $\rcrit$; $\cbary=1$ corresponds to my fiducial assumption that
baryons trace the total matter distribution.

The time-averaged star formation rate in this region is then 
\begin{flalign}
    \label{eq:sfr}
    \dot{M}_{\star}&=\epsff\,\frac{M_{\rm b}(<\rcrit)}{\tff(\rcrit)}\,\\
    &=\frac{\epsstar}{\etaff} \frac{M_{\rm b}(<\rcrit)}{\tff(\rcrit)}\,
\end{flalign}
i.e., a mass $\epsff\,\mbary(<\rcrit)$ of stars will form per free-fall time,
with the total star formation persisting for a period of $\etaff$ free-fall
times, producing a total stellar mass of
$\mstar=\epsstar\,\mbary(<\rcrit)=\epsstar\,\fbary\,c_{\rm b}\,\mtotcrit$ with
$\epsstar=\etaff\,\epsff$. Numerical and observational arguments point to
$\etaff \approx 3$, i.e., star formation will persist for approximately 3
free-fall times (e.g., \citealt{elmegreen2000, grudic2018, kim2018b,
  guszejnov2023}). The integrated star formation efficiency\footnote{Note the
  definition of $\epsstar$ differs from \citet{boylan-kolchin2023}: there, it
  was defined as the fraction of a halo's baryons that have been converted into
  stars, whereas here it is the fraction of baryons within the high acceleration
  region of a halo -- typically a small fraction of a halo's total baryonic
  content, as shown in the right panel of Fig.~\ref{fig:m_z_baryon} -- that have
  been converted into stars.} $\epsstar$ should be high ($\sim 0.5$) since these
systems are in the high acceleration regime where feedback is ineffective; the
efficiency per free-fall time $\epsff$ will be lower by a factor of
$\etaff$. The total duration of star formation is predicted to be at most
$\etaff\,\tff \approx 40\,{\rm Myr}$ for $\mbary(<\rcrit)=2\times 10^{9}\,\msun$
($\mstar = 10^{9}\,\msun$ for $\epsstar=0.5$); star formation indicators that
are sensitive to longer timescales will therefore underestimate the true
instantaneous star formation rate in these regions of efficient star formation.

Using Eq.~\ref{eq:tcrit}, the star formation rate can be written as
\begin{equation}
\label{eq:sfr_alt}
    \dot{M}_\star=24 \,\msun\,{\rm yr}^{-1}\,\left(\frac{\epsstar\,\cbary}{0.5}\right)^{1/4}\, \left(\frac{\etaff\,}{3}\right)^{-1}\,
    \left( \frac{\mstar}{10^{9}\,\msun} \right)^{3/4}\,,  
\end{equation}
yielding a specific star formation rate of $24\,{\rm Gyr}^{-1}$ at
$\mstar=10^{9}\,\msun$ for the fiducial values of star formation parameters and
a stellar mass dependence of $\dot{M}_\star/\mstar \propto \mstar^{-1/4}$. In
this regime, the stellar mass dependence of the specific star formation rate is
set simply by $(\tff\,\etaff)^{-1}$, which is the time it takes to convert gas
into stars in the region of high acceleration or (total) surface density.

Equation~\ref{eq:sfr_alt} can be expressed in terms of UV magnitudes as well;
assuming no attenuation, the result is 
\begin{flalign}
    \nonumber
    M_{\rm UV,un}=&-21.7-1.875\,\log_{10}\left( \frac{\mstar}{10^{9}\,\msun} \,\right) -0.625\,\log_{10}\left(\frac{\epsstar\,c_{\rm b}}{0.5}\right)\\
    &+2.5\, \log_{10} \left(\frac{\mathcal{K}_{\rm UV}}{\mathcal{K}_{\rm UV, 0}} \frac{\etaff}{3} \right)\,,\label{eq:uvmags}
\end{flalign}
where $\mathcal{K}_{\rm UV}$ is the conversion factor between specific luminosity and star formation rate (e.g., \citealt{kennicutt1998a}) and $\mathcal{K}_{\rm UV,0}=1.15\times10^{-28}\,\msun\,\yr\,{\rm erg}^{-1}\,{\rm s\,Hz}$ is the value for a \citet{salpeter1955} IMF; for a \citet{chabrier2003} IMF, $\mathcal{K}_{\rm UV}/\mathcal{K}_{\rm UV, 0} \approx 0.6$ \citep{madau2014}, making a given $\mstar$ brighter by $0.55\,{\rm mag}$ in the UV.

\begin{figure}
 \centering
 \includegraphics[width=\columnwidth]{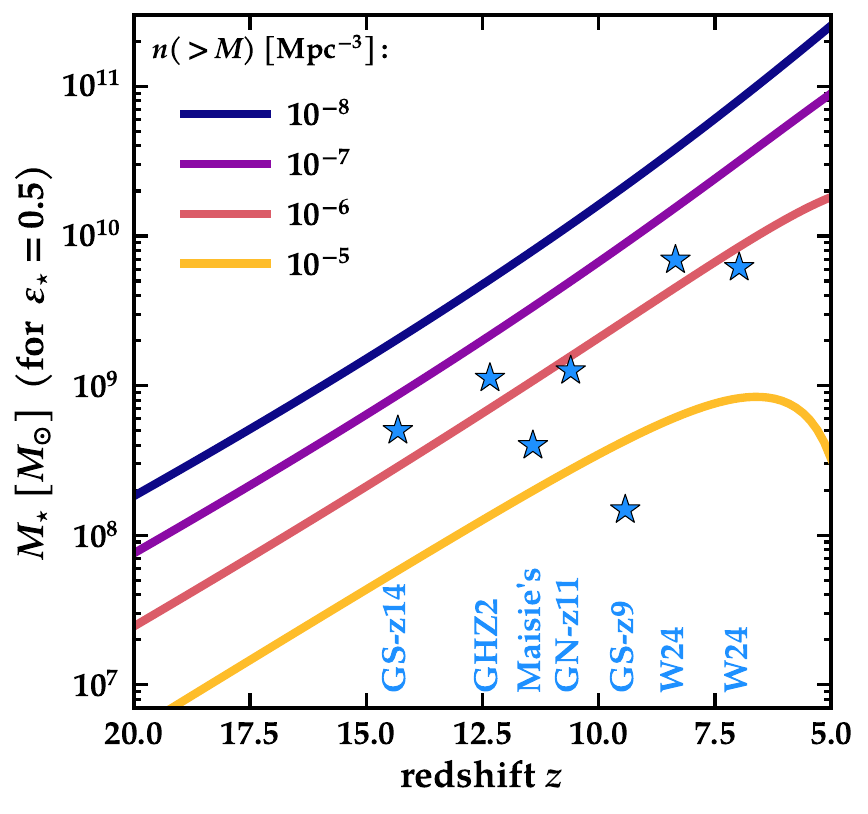}
 \caption{The curves give the expected evolution of stellar mass formed in
   regions of high acceleration as a function of $z$ at the fixed number
   densities listed on the plot. For comparison, several spectroscopically
   confirmed high-redshift galaxies are included as well; these galaxies are
   GS-z14 \citep{carniani2024}; GHZ12 \citep{castellano2022, naidu2022,
     castellano2024, zavala2025}; Maisie's galaxy \citep{finkelstein2022,
     arrabal-haro2023a}; GN-z11 \citep{oesch2016, bunker2023, tacchella2023};
   GS-z9 \citep{curti2024}; and RUBIES-EGS-55604/966323 (\citealt{labbe2023,
     wang2024a}; these are labeled as W24, and I have adopted the ``medium''
   stellar mass values from \citealt{wang2024a}). The implied number densities
   of these galaxies are consistent with the predictions from this paper for
   efficient dark-matter-driven galaxy formation at high accelerations, given
   the volumes surveyed by CEERS and JADES of $V\approx ({\rm a\,few})\times
   10^{5}\,\mpc^{3}$.
 \label{fig:mstar_z_obs}
}
\end{figure}

\subsection{Comparison with Observations}
\label{subsec:obs}
Figure~\ref{fig:mstar_z_obs} shows the evolution of $\mstar$ in systems with
fixed cumulative comoving number densities of
$n(>\mvir)=10^{-(5,\,6,\,7,\,{\rm and}\,8)}\,\mpc^{-3}$ as a function of
redshift. The figure assumes halos lie on the mean $c(\mvir|z)$ relation from
\citet{ishiyama2021} and that the stellar mass is formed in the region of high
acceleration as in the previous subsection, i.e.,
$\mstar=\epsstar\,\mbary(<\rcrit)$ with $\epsstar=\etaff\,\epsff$; for the
purposes of the figure, I assume an integrated star formation efficiency in this
phase of $\epsstar=0.5$. Several high-redshift galaxies with redshifts that have
been spectroscopically confirmed by JWST are also shown on the plot. These
galaxies have inferred stellar masses that are consistent with the predictions
of efficient galaxy formation in the high (dark matter) acceleration regime, as
they lie close to the volumes surveyed by CEERS \citep{finkelstein2023} and
JADES \citep{eisenstein2023} of
$V\approx ({\rm a\,few})\times 10^{5}\,\mpc^{3}$. Equations~\ref{eq:rcrit} and
\ref{eq:rhocrit} indicate that roughly
$0.1\,\msun\,\pc^{-3} \approx 16\,{\rm cm}^{-3}$ of baryons will reside in a
region of $\sim 2\,\kpc$ for halos with $\mtotcrit \approx 10^{10}\,\msun$ or
$\mbary(<\rcrit) \approx 2 \times 10^{9}\,\msun$ ($\mstar \approx 10^9\,\msun$
for $\epsstar=0.5$. If the baryons collapse by a factor of $\sim 10$, their
density will be $\approx 10^{4}\,{\rm cm}^{-3}$ in a region of $100\,\pc$; this
gives a rough estimate of the resulting size of the region that will undergo
intense star formation.

The numbers in the previous paragraph compare reasonably well with observations:
for example, \citet{tacchella2023} find that GN-z11 has a stellar mass of
$\approx 10^{9}\,\msun$ within a half-light radius of $64\,\pc$ and a star
formation rate of $\approx 20\,\msun\,\yr$. All of these values, as well as its
observed $M_{\rm UV}$ value of $-21.6\,{\rm mag}$ with $0.2\,{\rm mag}$ of
obscuration, agree very well with the expectations of the model described
here. From Figure~\ref{fig:m_z_baryon}, the expected halo mass is
$\mvir \approx 10^{10.7}-10^{11}\,\msun$, again consistent with the value
estimated by \citet{tacchella2023}. GN-z11 is the brightest of the galaxies in
Fig~\ref{fig:mstar_z_obs}, perhaps indicating it has been caught directly during
its maximally efficient star formation phase; many of the other galaxies may be
observed at a time somewhat offset from the maximum efficiency, reducing their
brightnesses and inferred star formation rates. GHZ12 has an effective radius of
$R_{\rm e} \approx 100\,\pc$, a stellar mass of
$\mstar=1.1\times 10^{9}\,\msun$, $M_{\rm UV}=-20.53$, and
$\dot{M}_{\star}=5\,\msun\,\yr$ \citep{castellano2024}. The predictions of this
work would put it $1-1.5$ magnitudes brighter and with a star formation rate
that is $3-5$ times higher if all of its stellar mass were formed in a single
burst of efficient star formation. If, instead, roughly 20\% of the mass was
formed very recently in the regime facilitated by high dark matter accelerations
--- or if star formation persisted for closer to $10\,\tff$ --- then the model
described here would provide a good match to the observations.

Based on Figures~\ref{fig:m_z_baryon} and \ref{fig:mstar_z_obs}, the mass in
baryons within the region of efficient star formation should drop by a factor of
$\approx 2$ from $z=14$ to $z=16$ and an additional factor of 2 at $z=18$ at a
fixed halo number density. The detection of GS-z14 \citep{carniani2024}
therefore augurs well for possible future confirmation of similar or slightly
less bright/massive galaxies out to $z\sim 18$ in the context of the
dark-matter-driven star formation model described here.

\section{Discussion}
\label{sec:speculation}
The idea at the heart of this paper is very simple: at fixed virial mass, dark
matter halos at high redshift are much denser than at low redshift, and this
higher density can lead to large quantities of baryons experiencing
accelerations high enough that stellar feedback should become ineffective. In
this regime, \textit{dark matter} is the source of high acceleration needed for
efficient star formation, which can occur on a scale much larger than at low
redshift, where such high accelerations are only realized in dense cores of
molecular clouds from the self-gravity of baryons. Some halos at high redshift
therefore should be able to form stars very efficiently owing simply to their
high densities. This general picture appears unavoidable.

The details of this process will depend on a number of factors, including the
concentrations of dark matter halos (which control the amount of mass above
$\gcrit$) and the value of $\gcrit$ itself. However, $\gcrit$ should not be
thought of as a threshold but rather as a characteristic value that roughly
separates inefficient star formation (at $g \ll \gcrit$) from highly efficient
star formation (at $g \gg \gcrit$; see, e.g., \citealt{grudic2020,
  hopkins2022}). The overall picture described here appears robust so long as
there is no mechanism for reducing the central densities of halos at very high
redshifts. One such candidate would be star formation feedback, but as
demonstrated in Figure~\ref{fig:gvir_func}, the halos in the high acceleration
regime enter this regime early enough that it appears implausible that they have
formed any significant amount of stars before efficient star formation
begins. If the stellar IMF differs substantially in bright systems at early
cosmic times relative to lower redshifts, $\gcrit$ could vary as well: for
example, a more top-heavy IMF, as has been invoked to explain JWST observations
\citep{inayoshi2022, steinhardt2023, lu2024, menon2024, van-dokkum2024}, would
increase $\langle \dot{p}/m_{\star}\rangle$ and therefore $\gcrit$. It would be
interesting to consider extensions of this work that take into account possible
variations in the IMF.

Modifying the assumed cold and collisionless nature of dark matter could also
affect the distribution of matter in the centers of dense dark matter halos: for
example, dark matter self-interactions tend to reduce the central densities of
halos \citep{spergel2000, bullock2017, buckley2018, tulin2018}. However, this
process is likely to be very inefficient in the regime considered here, where
halos are undergoing rapid mass assembly that serves as a heat supply,
preventing efficient core creation \citep{dave2001}. AGN feedback can provide
stronger outward accelerations than stellar feedback, but this requires massive
black holes; it is likely to operate only after an epoch of efficient galaxy
formation and black hole growth.

It is important to note that while the process discussed here is posited to lead
to high galaxy-wide star formation efficiency, the efficiency does \textit{not}
approach the theoretical maximum of $\fbary\,\mvir$ \citep{steinhardt2016,
  behroozi2018, boylan-kolchin2023}, as is demonstrated in the right panel of
Figure~\ref{fig:m_z_baryon}. This reflects the difficulty --- likely
impossibility --- of converting virtually all baryons in a halo into stars, as
most baryons reside at low densities far from the halo's center. Any significant
population of galaxies that require integrated star formation efficiencies of
$\mstar/(\fbary\,\mhalo) \approx 1$ would remain very difficult to understand
within \lcdm. Nevertheless, high acceleration from dark matter appears to be an
attractive and natural mechanism for explaining the surprisingly abundant and
bright galaxies in the infant cosmos revealed by JWST as well as why this
efficient star formation on large scales cannot continue to lower
redshifts. Failure of feedback in regions with sufficiently high accelerations
also dovetails naturally with the idea that the shorter dynamical timescales in
high-redshift systems many allow substantial periods of star formation prior to
the full onset of supernova feedback (e.g., \citealt{pallottini2023, li2023}).

From Figure~\ref{fig:gvir_func}, however, it is clear that many halos in the
efficient galaxy formation regime of $g>\gcrit$ will remain there for a
substantial period of time. This may lead to repeated cycles of efficient bursts
of star formation followed by (temporary) quiescence as gas re-accumulates at
the centers of halos. A rough estimate of the time-scales involved is that the
efficient bursts should occur on a local free-fall time (Eq.~\ref{eq:tcrit}),
which is very close to a local crossing time, while the resupply timescale is
comparable to the crossing time at the virial radius, which is
$t_{\rm vir}=75\,{\rm Myr}$ at $z=10$ independent of halo mass and scales as
$(1+z)^{-3/2}$ at high redshift. Rarer halos will have higher values of
$\tff/t_{\rm vir}$ at all epochs, meaning the duration of efficient starbursts
will be a larger fraction of the re-accretion time scale; as a result, very rare
and massive halos of $n(>\mvir)\approx 10^{-8}\,\mpc^{-3}$ may go through cycles
of efficient bursts with a duty cycle of $\sim 35\%$, whereas halos with
$n(>\mvir)\approx 10^{-5}\,\mpc^{-3}$ will have duty cycles of at most
$15\%$. The star formation histories of the most massive and rarest galaxies
therefore may show evidence of more continuous efficient star formation than
more typical galaxies.

As shown in Figure~\ref{fig:m_vs_z} and discussed in Sec.~\ref{subsec:z_dep},
the redshift dependence of the threshold mass for efficient star formation in my
model is eerily close to the threshold for feedback-free bursts posited by
\citet{dekel2023} and discussed further in \citet{li2023}. While some
similarities certainly exist, the two models rely on very different assumptions
and make substantively different predictions: for example, \citet{dekel2023}
quote an expected stellar mass of $\mstar \approx 10^{10}\,\msun$ at
$z\approx 10$ in halos of $\mvir \approx 10^{10.8}\,\msun$ with a star formation
rate of $65\,\msun\,\yr$; at the same halo mass, the model described here would
result in an order of magnitude lower stellar mass (see
Fig.~\ref{fig:m_z_baryon}) as well as a star formation rate that is lower by a
factor of $2-3$. An avenue of future interest is a more detailed comparison of
the two models and an exploration of whether their predictions are in conflict
or concordance.

At the high accelerations considered here, the high efficiency of star formation
is not the only expected change: stars should form preferentially in self-bound
clusters \citep{hills1980, krumholz2019, li2019}, with \citet{grudic2021}
finding that the fraction of stars forming in bound clusters approaches unity at
integrated star formation efficiencies of in excess of $\epsstar \approx
0.25$. The result of the star formation process postulated here should therefore
be a region of a galaxy dominated by young star clusters. This scenario is
supported by JWST observations that have revealed lensed systems with a large
number of infant clusters dominating the light (e.g., \citealt{adamo2024a,
  bradley2024, fujimoto2024}. Figure~\ref{fig:m_z_baryon} indicates that
sufficient collections of baryons ($\approx 10^{6}\,\msun)$ may be subjected to
high enough accelerations to form individual globular clusters as early as
$z \approx 20$ in halos with volume densities of
$10^{-4}\,\mpc^{-3}$. Furthermore, the mass in baryons above $g_{\rm crit}$ in
the lowest-mass (and therefore most common) halos achieving $g_{\rm thresh}$ at
their centers is $\sim 10^{6}\,\msun$, a mass scale intriguingly similar to that
of globular clusters, at all redshifts.

Massive, dense clusters are potentially the sites of top-heavy IMFs (e.g.,
\citealt{haghi2020}) and may host supermassive stars (a leading candidate to
explain anomalous chemical abundances observed in massive globular clusters in
the Milky Way; \citealt{denissenkov2014, bastian2018}), which means the mode of
star formation proposed here might be conducive to the formation of IMF
variations and massive black hole seeds (the remnants of the supermassive
stars). An additional change relative to standard theories of star formation at
high surface density in the dark-matter-driven high efficiency regime described
here is that escape velocity from star-forming regions will be much higher than
for typical molecular clouds owing both to their greater masses and to the large
reservoirs of dark matter on somewhat larger scales. This may result in more
efficient self-enrichment of galaxies --- and possibly even star clusters ---
formed in this way at early cosmic epochs, potentially imprinting a signature of
this mode of efficient galaxy formation and helping to explain abundance
anomalies observed in a subset of stars in massive globular clusters. Another
intriguing possibility is that the mechanism discussed here could be conducive
to the formation of direct collapse black holes in metal-free gas at higher
redshift. Figure~\ref{fig:m_z_baryon} hints that this may be possible.

One final point of interest relates to the evolution of densities under
hierarchical assembly. The high densities at high redshifts described here occur
within a fixed physical radius; the fact that similar densities do not typically
occur in more massive halos at lower redshifts indicates that either (1) there
must be a mechanism for reducing dark matter densities through hierarchical
assembly, or (2) the descendants of these halos with high accelerations at high
redshift survive to the present day with similarly high physical densities at
their centers.

Option (1) would be somewhat surprising, as controlled simulations of dark
matter halo mergers indicate that central densities increase in physical units
as a result of the merger process \citep{boylan-kolchin2004, kazantzidis2006,
  drakos2019}. \citet{diemand2007a} also demonstrate the relative constancy of
$M(<r)$ at small radii within fixed physical apertures for the cosmological
evolution of an individual Milky-Way-mass halo. However, the required effect
need only operate in rare halos with number densities less than
$\sim 10^{-5}\,\mpc^{-3}$; it is not surprising that it has not been observed in
zoom-in cosmological simulations focusing on Milky-Way-mass systems, which are
substantially more common. Zoom-in simulations of dark matter halos at the scale
of galaxy clusters (e.g., \citealt{gao2012a}) are therefore of great interest in
this context. Moreover, the models of \citet{loeb2003} and \citet{gao2004}
\textit{do} point to a reduction of physical densities in the centers of halos
with fixed $n(>\mvir)$ over cosmic time; investigating possible mechanisms for
such a reduction is an important avenue for future work.

Option (2) could be realized if the high acceleration halos end up with high
concentrations for their mass at $z=0$, as the mass within fixed physical
apertures in the inner regions of such halos will be larger than for typical
halos, or as dense substructure in more massive systems (see also
\citealt{ishiyama2014, errani2018, van-den-bosch2018, delos2023}). This
possibility is intriguing, as it points to galaxies living in the
earliest-forming massive halos as excellent sites for probing efficient galaxy
formation at high redshift. Indeed, the physical \textit{stellar} densities of
the most massive galaxies at high redshift are comparable to those of the most
massive ellipticals in the local Universe (e.g., \citealt{hopkins2010,
  baggen2023}). A more detailed analysis that folds in the full distribution of
halo concentrations at fixed virial mass and its evolution with time would be
highly valuable in evaluating whether option (2) is a viable explanation.

\section{Conclusions}
\label{sec:conclusions}
While star formation is generally inefficient when considered as the fraction of
gas turned into stars on a local dynamical time or integrated over the lifetime
of a star-forming region, efficient star formation \textit{can} happen when
stellar feedback cannot overcome the gravity of star-forming gas. This regime is
characterized by acceleration that exceed
$\gcrit \approx 5\times 10^{-10}\,\mss$ (or
$\sigcrit=\gcrit/(\pi\,G) \approx 1000\,\mspsq$), which is set by the momentum
flux per unit mass $\langle \dot{p}/m_{\star}\rangle$ from a young stellar
population. In the low-redshift Universe, the only regions where such
accelerations are realized --- dense clumps within molecular clouds --- are
baryon-dominated. However, I point out in this paper that at high redshift, the
significantly higher mean density of the Universe results in regions within
galaxy-mass halos where dark matter can provide the necessary accelerations for
efficient formation of galaxy-scale quantities of stars
($\mstar \sim 10^{8}-10^{10}\,\msun$).

This straightforward but surprising result has important implications for our
understanding of galaxy formation at high redshifts ($z \gtrsim 8)$, where JWST
has revealed unexpectedly bright and massive galaxies. The basic picture I
describe in this paper can be summarized as follows. The virial mass
corresponding to a fixed virial acceleration scales as $(1+z)^{-6}$
(Eq.~\ref{eq:mvir_gvir}), tracking roughly constant cumulative comoving number
densities of halos at early times. Assuming that dark matter halos have NFW
profiles with concentrations that follow the mean relations measured in
cosmological simulations, there is a threshold virial acceleration of
$\gvir/G \approx 380\,\mspsq$: above this value, the central portion of the halo
will experience accelerations in excess of $\gcrit$
(Figure~\ref{fig:m_vs_z}). This threshold virial acceleration corresponds to
$n(>\mvir) \approx 10^{-4}\,\mpc^{-3}$. The amount of baryonic mass contained in
the region of high acceleration is $\approx 10^{6}\,\msun$ at the threshold
mass; for more massive (and therefore rarer) halos, the mass in baryons subject
to high accelerations can be comparable to the observed masses of the highest
redshift galaxies (Figure~\ref{fig:m_z_baryon} and \ref{fig:mstar_z_obs}).

The regions of where dark matter provides acceleration in excess of $\gcrit$ are
characterized by initial sizes of $\approx 2\,{\rm kpc}$, baryonic densities of
$\approx 0.1\,\msun\,\pc^{-3}$, and free-fall times of $\approx 13\,{\rm Myr}$
for baryonic content of $\mbary(>\gcrit) \approx 2 \times 10^{9}\,\msun$
(Equations~\ref{eq:rcrit}-\ref{eq:tcrit}). Assuming an integrated star formation
efficiency of $\epsstar=0.5$ in this region results in a stellar mass of
$10^{9}\,\msun$ that will be formed in $\approx 40\,{\rm Myr}$, a star formation
rate of $24\,\msun\,\yr$ over this period, and an unattenuated UV magnitude of
$-21.7$ (assuming a Salpeter IMF). The specific star formation rate is expected
to scale as $\mstar^{-1/4}$. To reach densities required for star formation, the
baryons in a region such as this must collapse by a factor of $\approx 10$,
giving a size of $\mathcal{O}(100\,\pc)$. These properties are in reasonable
agreement with observations of luminous systems at high redshift such as GN-z11
(Sec.~\ref{subsec:obs}), with the prediction that they reside in halos with
number densities of $n(>\mvir) \approx 10^{-5.5}-10^{-6.5}\,\mpc^{-3}$
(Figure~\ref{fig:mstar_z_obs}).

The dependence of the threshold virial acceleration on redshift is nearly
identical to what was predicted for feedback-free bursts in \citet{dekel2023},
an intriguing similarity given the differences in the underlying physical
models. In detail, the predictions here differ non-trivially from those for
feedback-free bursts, with dark-matter-driven efficient galaxy formation
predicting lower global star formation efficiencies and lower stellar masses at
fixed halo mass. Future avenues for exploration include folding in a full
$\epsstar-g$ relation as described in, e.g., \citet{fall2010},
\citet{grudic2018}, or \citet{hopkins2022} and a cosmological distribution of
concentrations at fixed halo mass and redshift (as the central gravitational
acceleration at a given halo mass and redshift depends only on concentration via
Eq.~\ref{eq:gmax}). Understanding the fate of the predicted regions of high
galaxy formation efficiency will also be important, as within the basic paradigm
described in this paper, they must either become less dense with time or
represent the high-concentration tail of massive halos (or their substructure)
in the local Universe, as described at the end of
Section~\ref{sec:speculation}. Nevertheless, the simplicity and predictive power
of the model presented here for efficient dark-matter-driven star formation on
galactic scales make it a promising explanation for the highly active earliest
epochs of galaxy formation revealed by JWST.

\section*{Acknowledgments} 
I thank Volker Bromm, James Bullock, Neal Evans, Phil Hopkins, Pawan Kumar,
Stella Offner, Eliot Quataert, Jenna Samuel, and Stuart Wyithe for helpful
conversations that informed this work. I acknowledge support from NSF CAREER
award AST-1752913, NSF grants AST-1910346 and AST-2108962, NASA grant
80NSSC22K0827, and HST-GO-16686, HST-AR-17028, and HST-AR-17043 from the Space
Telescope Science Institute, which is operated by AURA, Inc., under NASA
contract NAS5-26555. I am very grateful to the developers of the python packages
that I used in preparing this paper: {\sc numpy} \citep{numpy2020}, {\sc scipy}
\citep{scipy2020}, {\sc matplotlib} \citep{matplotlib}, {\sc ipython}
\citep{ipython}, {\sc hmf} \citep{murray2013, murray2014}, and {\sc colossus}
\citep{diemer2018}. This research has made extensive use of NASA’s Astrophysics
Data System (\url{http://adsabs.harvard.edu/}) and the arXiv e-Print service
(\url{http://arxiv.org}).

\section*{Data availability}
No new data were generated or analyzed in support of this research.

\bibliography{draft_clean}
\end{document}

%% file: extra_preamble.tex
\newcommand{\mvir}{M_{\rm{vir}}}

\newcommand{\rvir}{R_{\rm{vir}}}

\newcommand{\mstar}{M_{\star}}
\newcommand{\msun}{M_{\odot}}

\newcommand{\mpc}{{\rm Mpc}}
\newcommand{\kpc}{{\rm kpc}}

\newcommand{\pc}{{\rm pc}}
\newcommand{\kms}{{\rm km \, s}^{-1}}

\newcommand{\lcdm}{$\Lambda$CDM}

\newcommand{\yr}{{\rm yr^{-1}}}

\newcommand{\mhalo}{M_{\rm halo}}

\newcommand{\ssim}{{\sim}}
\renewcommand{\ga}{\gtrsim}
\renewcommand{\la}{\lesssim}